\documentclass[prl,aps,onecolumn,11pt]{revtex4-2}
\usepackage{graphicx}
\usepackage{xcolor}

\begin{document}
\title{Continuous Acceleration Sensing Using Optomechanical Droplets}

\author{G.R.M. Robb, J.G.M. Walker, G.-L. Oppo \& T. Ackemann}
\affiliation{SUPA and Department of Physics, University of Strathclyde,Glasgow G4 0NG, Scotland, UK}


\begin{abstract}
We show that a Bose--Einstein Condensate illuminated by a far off-resonant optical pump field and its retroreflection from a feedback mirror can produce stable, localised structures known as  optomechanical droplets. We show that these droplets could be used to measure the acceleration of a BEC via continuous monitoring of the position of the {droplet via} the optical intensity distribution. 
\end{abstract}

\maketitle

\section{Introduction}
The utility of cold atomic gases and Bose--Einstein Condensates (BECs) for applications involving sensing of various quantities (from acceleration  to electromagnetic fields) has been recognised for many years now~\cite{barrett2014,kasevich1992,adams1994,peters1999,bloom1962,budker2007,kitching2011}. In~acceleration sensing involving ultracold atoms or BECs, many traditional measurement techniques such as time-of-flight measurements are destructive, with~each measurement requiring a new sample of atoms~\cite{peters1999}. Dispersive imaging techniques which utilise the refractive properties of the atoms have been used~\cite{andrews1996,andrews1997,bradley1997,stenger1999,saba2005} to realise non-destructive imaging of~BECs. 

Recently, there has been interest in continuous, minimally destructive sensing of  acceleration via interaction of a BEC with light contained within an optical cavity~\cite{peden2009,venkatesh2009,goldwin2014,kessler2016,samoylova2015_1,samoylova2015_2}. The~dynamical behaviour of the BEC under the action of an external force, e.g.,~gravity, influences the dynamical behaviour of the optical field in the cavity; the~BEC dynamics can be inferred from measurement of the optical field escaping from the cavity and the acceleration of the BEC can be calculated. This concept has been considered in several different configurations involving both Fabry--Perot and ring cavities where the BEC exhibits Bloch oscillations due to its acceleration through a spatially periodic optical lattice~\cite{salomon1996} consisting of counterpropagating cavity modes alone~\cite{peden2009,venkatesh2009,goldwin2014,kessler2016} or by adding an externally applied optical lattice~\cite{samoylova2015_1,samoylova2015_2}.

\begin{figure}
\includegraphics[width=0.7\columnwidth]{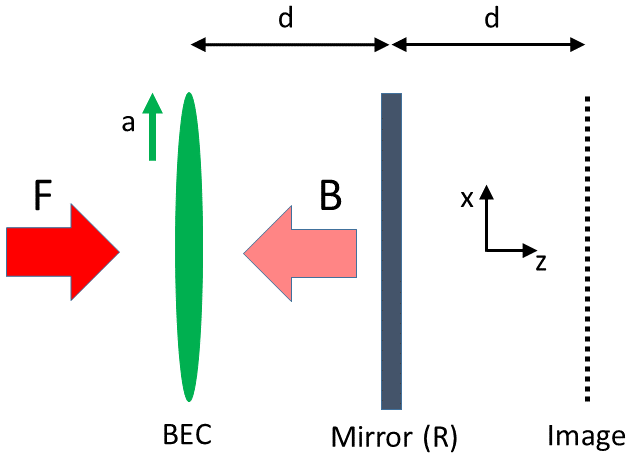}
\caption{Schematic diagram of the single mirror feedback (SMF) configuration showing a BEC interacting with a forward propagating optical field ($F$) and a retroreflected/backward propagating optical field ($B$) while undergoing an acceleration ($a$) in the $x-$direction. The~optical image of the BEC is detected after transmission through the BEC, propagation in free space to a mirror of reflectivity, $R$, at~a distance $d$ from the BEC,  and~further propagation over distance $d$.}
\label{fig:schematic}
\end{figure}

The interaction between cold atoms/BECs and light in an optical cavity has also attracted significant interest due to the existence of self-organisation phenomena, e.g.,~self-organised optical/atomic patterns~\cite{Kruse2003CARL,Ritsch2013cavities,Mivehar2021cavities,kollar17, kroeze2018spinor, guo21a}. {Self-organized structures and patterns  can also be produced by the simultaneous presence of optical nonlinearity and diffraction. These patterns have been predicted and observed in a variety of nonlinear media~\mbox{\cite{cross1993pattern, grynberg1988observation, lippi1994transverse1, lippi1994transverse2, giusfredi1988optical, firth1990spatial, arecchi99, lugiato94b, rosanov96, barbay11}} specifically including  atomic vapours~\cite{cross1993pattern, grynberg1988observation, lippi1994transverse1, lippi1994transverse2, giusfredi1988optical, firth1990spatial, grynberg94, ackemann94, ackemann01}. In~the case of cold atoms/BECs, the origin of optical nonlinearity is the spatial modulation of atomic density which arises due to the mechanical effect of light, specifically  optical dipole forces. Formation of a spatially modulated atomic density and its subsequent backaction on  light give rise to an optomechanical self-structuring instability, resulting in the simultaneous, spontaneous formation of atomic density structures and optical intensity structures.  Optomechanical self-structuring of a cold thermal gas has been studied experimentally and theoretically in systems of counterpropagating beams~\cite{greenberg11, schmittberger16} and in a single mirror feedback (SMF) configuration~\mbox{\cite{labeyrie2014optomechanical, tesio2014theory,firth2017}}. Theoretical predictions of optomechanical self-structuring in a BEC~\cite{robb2015} highlighted that a significant difference with the classical, thermal gas is the the dispersive nature of the BEC wavefunction, i.e.,~``quantum pressure'', which acts to suppress density modulations or spatial structures in the BEC. 
Recent work has shown that optomechanical self-structuring in a BEC can produce spatially localised structures termed ``droplets'' or ``quantum droplets''~\cite{zhang2018long, zhang2021self,walker2022} in addition to global patterns.  These droplets are self-bound  structures consisting of interacting light and matter whose stability is reliant upon the BEC quantum pressure. They display some similar characteristics to quantum droplets in other systems such as dipolar BECs~\cite{edmonds2020quantum} and quantum liquids~\cite{cabrera2018} but also have some properties similar to those of other varieties of spatially localised structures, e.g.,~spatial solitons~\cite{ackemann2009fundamentals}.
In this paper, we study the dynamical behaviour of these optomechanical droplets in a 1D configuration involving a single feedback mirror, as~in~\cite{zhang2018long,zhang2021self}, and~consider the effect of a uniform BEC acceleration, shown schematically in Figure~\ref{fig:schematic}. }

\section{Model}
{We investigate the system shown schematically in Figure~\ref{fig:schematic} consisting of a BEC with single mirror feedback (SMF). In~this BEC--SMF system, coupling between atoms arises due to their interaction with a pump optical field and its reflection from the feedback mirror. Diffraction of the optical field as it propagates from the BEC to the mirror and back again plays a critical role in this coupling. The~interaction involves many transverse modes and optical forces directed perpendicular to the propagation direction of the optical fields. A significant difference between this system and other cavity systems displaying self-organization (such as e.g. \cite{voncube2004,baumann2010dicke}) is that interference between the optical fields does not play a significant role in this system, whereas in.~\cite{voncube2004,baumann2010dicke} the dominant coupling between atoms arises from interference between a pump field and cavity modes. We demonstrate the existence of spatially localised states with characteristics similar to those of quantum droplets observed in dipolar BECs~\cite{ferrier2016,chomaz2016} and the effect of uniform acceleration on these~droplets. }

The model we use to describe the BEC--SMF system was originally studied in~\cite{robb2015}. {We consider a BEC with a negligible scattering length and describe  the evolution of the BEC wavefunction, $\Psi(x,t)$, with the Schr{\"o}dinger equation:}
\begin{equation}
\label{eqn:sch}
i \frac{\partial \Psi(\bar{x},\bar{t})}{\partial \bar{t}} = - \bar{\omega}_r \frac{\partial ^2 \Psi}{\partial \bar{x}^2} + \frac{\Delta}{4} \left(|F|^2 + |B|^2 \right) \Psi
\end{equation}
where $\bar{t}= \Gamma t$ and $\bar{x} = q_c x$ are dimensionless time and space variables, $q_c$ is a critical wavenumber to be defined shortly, $\Gamma$ is the decay rate of the atomic transition and $\bar{\omega}_r = \frac{\hbar q_c^2}{2 m \Gamma}$ is a dimensionless recoil frequency, where $m$ is the atomic mass.  {The quantities $|F|^2$ and $|B(x,t)|^2$ are the atomic saturation parameters due to the forward and backward optical fields defined as  $|F|^2 = \frac{I_F}{I_{sat} \Delta^2}$  and $|B|^2 = \frac{I_B}{I_{sat} \Delta^2}$, respectively. $\Delta = \frac{2 \delta}{\Gamma}$ is a dimensionless detuning parameter where $\delta = \omega - \omega_a$ is the detuning between the optical field frequency, $\omega$, and~the atomic transition frequency, $\omega_a$, and $I_F$ and $I_B$ are the intensities of the forward ($F$) and backward ($B$) fields, respectively. $I_{sat}$ is the saturation intensity on resonance. 
It has been assumed that the optical fields are far-detuned from atomic resonance ($|\Delta| \gg 1$)  and that consequently $|F|^2,|B|^2 \ll 1$  so that the atoms remain in their ground state.  In~addition, any effects of gratings formed along the propagation ($z$) axis due  to  interference  between  the counterpropagating optical fields are neglected.}

{In order to describe the evolution of the optical field in the BEC, we assume that the gas is diffractively thin; i.e., it~is sufficiently thin that diffraction can be neglected. Consequently, the~forward field transmitted through the cloud is}
\begin{equation}
F_{tr}= \sqrt{p_0} e^{-i \chi_0 n(x,t)}
\label{eqn:Ftr}
\end{equation}
where $p_0 = |F(z=0)|^2$ is the scaled pump intensity, $\chi_0 = \frac{b_0}{2 \Delta}$ is the susceptibility of the BEC, $b_0$ is the optical thickness of the BEC at resonance and $n(x,t)= |\Psi(x,t)|^2$ is the local BEC density, which for a BEC of uniform density is $n(x,t) = 1$.

{After propagation of this transmitted forward field from the BEC to the feedback mirror in free space, the~reflected backward field, $B$, at~the BEC completes the feedback loop. As~the field propagates a distance $2d$ from the BEC to the  mirror  and  back,  optical phase modulations induced by transmission through density modulations in the BEC are converted to optical amplitude modulations and consequently optical dipole forces.  The~Fourier components of the forward and backward fields at the BEC are related by }
\begin{equation}
B(q) = \sqrt{R} F_{tr}(q) e^{i \frac{q^2 d}{k_0}}
\label{eqn:B}
\end{equation} 
where $R$ is the mirror reflectivity, $q$ is the transverse wavenumber and $k_0 = \frac{2 \pi}{\lambda_0}$.
It was shown in~\cite{robb2015} that this system exhibits a self-structuring instability, where the optical fields and BEC density develop modulations with a spatial period of $\Lambda_c = \frac{2 \pi}{q_c}$, where the {critical} wavenumber, $q_c$, is
\begin{equation}
q_c = \sqrt{\frac{\pi}{2} \frac{k_0}{d}}.
\label{eqn:qc}
\end{equation}

It can be seen from Equation~(\ref{eqn:qc}) that $\Lambda_c$ can be varied by changing $d$, the~mirror distance from the atomic cloud. {Typical values of $\Lambda_c$  used for thermal gases are 50--150 $\mu$m and we anticipate 5--20 $\mu$m to be suitable for a BEC experiment, corresponding to mirror distances in the range of  $d \approx$  8--150 $\mu$m. As~typical BEC thicknesses  are in the range of 1--10 $\mu$m, we expect from  previous investigations that the ``thin diffractive medium'' approximation still holds qualitatively and certainly gives a good initial indication useful for first demonstration of the principle discussed here. Ref.~\cite{firth2017} describes a more rigorous theoretical model for thermal atoms, taking diffraction within the medium into account.}

{The origin of the instability is BEC density modulations (which create refractive index modulations) with spatial frequency $q_c$. These density modulations in turn produce optical phase modulations in $F_{tr}$, which then are converted into optical intensity modulations of the reflected field $B$ at the BEC (see Equation~(\ref{eqn:B})). These optical intensity modulations result in dipole forces which act to reinforce the BEC density modulations, providing positive feedback and consequently instability of the initial, homogeneous state. 
In order to realise this instability, the~pump intensity must exceed a threshold value, $p_{\rm th}$ \cite{robb2015}, which for $q=q_c$ can be written as }
\begin{equation}
p_{\rm th} = \frac{2 \omega_r }{b_0 R \Gamma} ,
\end{equation}
where $\omega_r=\frac{\hbar q_c^2}{2m}$.

\section{Existence of Optomechanical~Droplets}

\begin{figure}
\includegraphics[width=0.99\columnwidth]{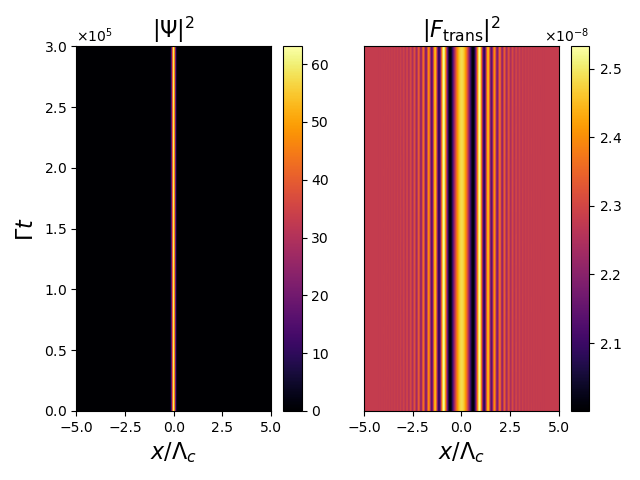}
\caption{Evolution of the BEC density, $|\Psi(\bar{x},\bar{t})|^2$,  and~optical intensity at the image plane, $|F_{\rm trans}(\bar{x},\bar{t})|^2$, calculated from the BEC--SMF model, Equations~(\ref{eqn:sch})--(\ref{eqn:B}), using an initial condition where the BEC density is a Gaussian function of position with width $\sigma_{\bar{x}} = 0.562$. Parameters used are $\bar{\omega}_r = 1.14 \times 10^{-5}$, $b_0 = 100$, $\Delta=-10000$, $R=0.99$ and $p_0 = 10 p_{\rm th} = 2.28 \times 10^{-6}$.} 
\label{fig:droplet_stationary}
\end{figure}

Numerical simulations of the BEC--SMF model, Equations~(\ref{eqn:sch})--(\ref{eqn:B}), using an initial condition where the BEC density is initially a Gaussian function of position, i.e.,
\begin{equation}
\label{eqn:Gaussian}
n(\bar{x},\bar{t}=0) = |\Psi(\bar{x},\bar{t}=0)|^2 \propto \exp \left( -\frac{\bar{x}^2}{\sigma_{\bar{x}}^2}\right)
\end{equation}
show that for certain BEC widths, $\sigma_{\bar{x}}$, the~BEC profile remains constant as time evolves, as~shown in Figure~\ref{fig:droplet_stationary}.

This behaviour is consistent with simulations involving integration of the BEC--SMF model, Equations~(\ref{eqn:sch})--(\ref{eqn:B}), in imaginary time~\cite{goldberg1967}, which show that the ground state of the BEC + optical field system is a localised droplet~state. 

It was shown in~\cite{robb2023} that it is possible to map the BEC--SMF model, Equations~(\ref{eqn:sch})--(\ref{eqn:B}), onto the quantum Hamiltonian Mean Field (quantum HMF) model, a~Gross--Pitaevskii-like equation for the BEC wavefunction, $\Psi(x,t)$, involving a non-local potential. The~quantum HMF model is known to possess soliton solutions~\cite{plestid2019} which correspond to the localised optomechanical droplets which occur in the BEC--SMF simulations (Figure~\ref{fig:droplet_stationary}). They can be understood as BEC gap solitons in an optical lattice which is self-generated by the BEC~\cite{plestid2019}. 

The width of a stable droplet can be calculated by assuming a Gaussian BEC density profile as in Equation~(\ref{eqn:Gaussian}) and minimising the energy, $E(\sigma_{\bar{x}})$, of~the system, resulting in~\cite{robb2023}
\begin{equation}
\label{eqn:droplet_width}
\sigma_{\bar{x}} = \left( \frac{p_0}{p_{\rm th}} \right)^{-1/4} 
\end{equation}
where $p_0 \gg p_{\rm th}$ has been assumed. From~the parameters used for Figure~\ref{fig:droplet_stationary}, the~predicted stable droplet width from Equation~(\ref{eqn:droplet_width}) is $\sigma_{\bar{x}} = 0.562$, which agrees well with the value calculated from simulations shown in Figure~\ref{fig:droplet_stationary}, which shows that a droplet of this width (which in the figure corresponds to a width $\sigma_x / \Lambda_c = \sigma_{\bar{x}}/(2 \pi) =  0.089$) is a stable, stationary~solution.

\section{Continuous Acceleration Sensing Using Optomechanical~Droplets}
\begin{figure}
		\includegraphics[width=1.0\columnwidth]{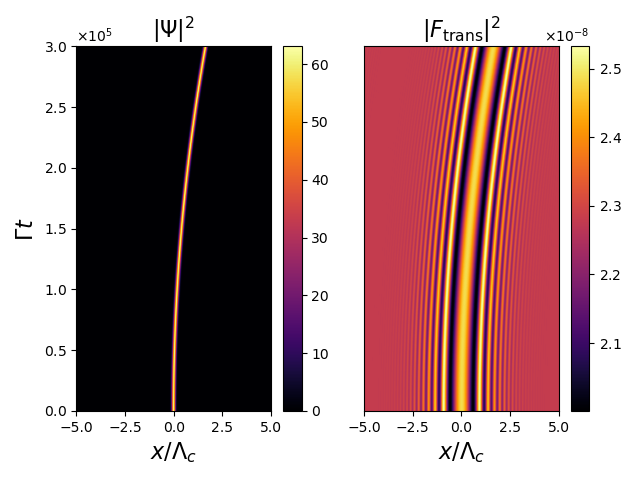}
		\caption{Evolution of the BEC density, $|\Psi(\bar{x},\bar{t})|^2$,  and~optical intensity at the image plane, $|F_{\rm trans}(\bar{x},\bar{t})|^2$, calculated from the accelerating BEC--SMF model, Equation~(\ref{eqn:Ftr}), (\ref{eqn:B}) and (\ref{eqn:sch2}), showing a uniformly accelerating droplet. Parameters used are identical those in Figure~\ref{fig:droplet_stationary} with the exception of the acceleration parameter, which here is {$\bar{a} = 1.0 \times 10^{-5}$}.}
		\label{fig:accelerating_red_droplet}
	\end{figure}

We now investigate the behaviour of these droplets under uniform acceleration. Modifying the Schr{\"o}dinger equation, Equation~(\ref{eqn:sch}), to~include uniform acceleration results~in
\begin{equation}
\label{eqn:sch2}
i \frac{\partial \Psi(\bar{x},\bar{t})}{\partial \bar{t}} = - \bar{\omega}_r \frac{\partial ^2 \Psi}{\partial \bar{x}^2} + \left[ \frac{\Delta}{4} \left(|F|^2 + |B|^2 \right) - \bar{a} \bar{x}  \right] \Psi
\end{equation}
where  $\bar{a} = \frac{m a}{\hbar q_c \Gamma}$ is a dimensionless acceleration parameter~\cite{peden2009,venkatesh2009,goldwin2014,kessler2016,samoylova2015_1,
samoylova2015_2,walker2022}.
{Figure~\ref{fig:accelerating_red_droplet} shows the evolution of the BEC density and the optical fields. Uniform acceleration of BEC can be observed, and~the optical field follows this motion, with the BEC density coinciding with an optical intensity maximum. }

{Figure~\ref{fig:accelerating_red_droplet} shows that via continuous observations of the optical intensity distribution, it is possible to infer the dynamical evolution of the BEC density.}

In order to perform a quantitative calculation of the uniform acceleration experienced by the BEC, we can use the fact that the position of an object undergoing uniform acceleration is described by
\[
\bar{x} = \bar{\omega}_r \bar{a} \bar{t}^2 
\]
so consequently
\begin{equation}
\label{eqn:trajectory}
\frac{x}{\Lambda_c} = \frac{\bar{\omega}_r \bar{a}}{2 \pi} (\Gamma t)^2 
\end{equation}

\begin{figure}
	\includegraphics[width=0.85\columnwidth]{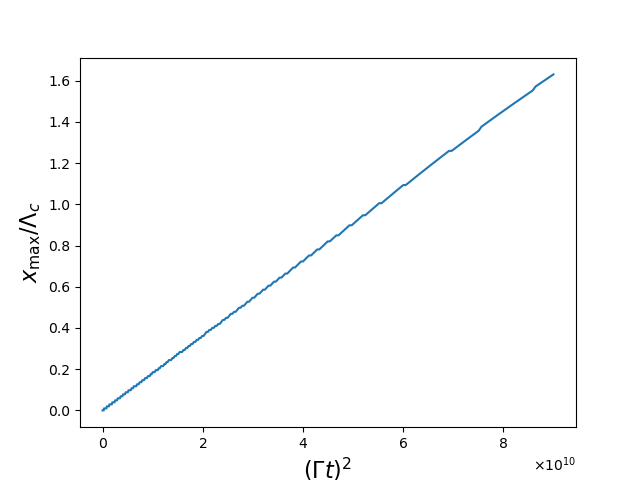}
	\caption{Plot of position of central optical intensity maximum, $\bar{x}_{\rm max}$, against~$\bar{t}^2$ from Figure~\ref{fig:accelerating_red_droplet}.}
	\label{fig:x_vs_tsq}
\end{figure}

Figure~\ref{fig:x_vs_tsq} shows a plot of the position of the central peak of the optical intensity in Figure~\ref{fig:accelerating_red_droplet} against $(\Gamma t)^2$. It can be seen that the graph is a straight line with gradient \mbox{$1.81 \times 10^{-11}$}. Equation~(\ref{eqn:trajectory}) implies that the gradient of this graph should be $\frac{\bar{\omega}_r \bar{a}}{2 \pi}$, so consequently $\bar{a} = \frac{2 \pi \times 1.81 \times 10^{-11}} {1.14 \times 10^{-5}} = 1.00 \times 10^{-5}$, consistent with the value of $\bar{a}$ used to produce Figure~\ref{fig:accelerating_red_droplet}. Continuous monitoring of the position of the optical intensity maximum has therefore been used to calculate the constant acceleration experienced by the BEC. Note that although the examples presented here  involved red-detuning ($\Delta < 0$), the~same procedure could have been used for blue-detuning ($\Delta > 0$). The~only significant difference would be that the BEC overlaps with an optical intensity minimum, but~again from monitoring the position of this, the~BEC acceleration could be~inferred. 

{The example shown in Figures~\ref{fig:accelerating_red_droplet} and \ref{fig:x_vs_tsq} is consistent with a Cs BEC droplet of width $\approx$0.5 $\mu$m moving a distance $\approx$10 $\mu$m in a time $\approx$0.01s, with~an acceleration $\approx$0.2 ms$^{-2}$. For~accelerations much smaller than this, a~limiting factor will be the heating of the BEC due to spontaneous light scattering. This will become significant when the interaction time becomes significantly larger than $r_s^{-1}$, where $r_s$ is the rate at which photons are incoherently scattered by the BEC. In~\cite{robb2023}, it is shown that
\[
r_s = \frac{(1+R) p_0 \Gamma}{2}
\]
so heating will be significant when the interaction time is sufficiently long that  $\Gamma t > \frac{2}{(1+R) p_0} \approx \frac{1}{p_0}$ (if $R \approx 1$). In~the example shown here, $\Gamma t = 3 \times 10^5$ and $\frac{1}{p_0} \approx 4.4 \times 10^5$, so heating will not have been significant. For~smaller accelerations, longer interaction times will be required in order to observe significant droplet displacement, so the effect of heating will be increasingly important. Conversely, larger accelerations will require shorter interaction times and heating will become negligible.}

\begin{figure}
		\includegraphics[width=0.99\columnwidth]{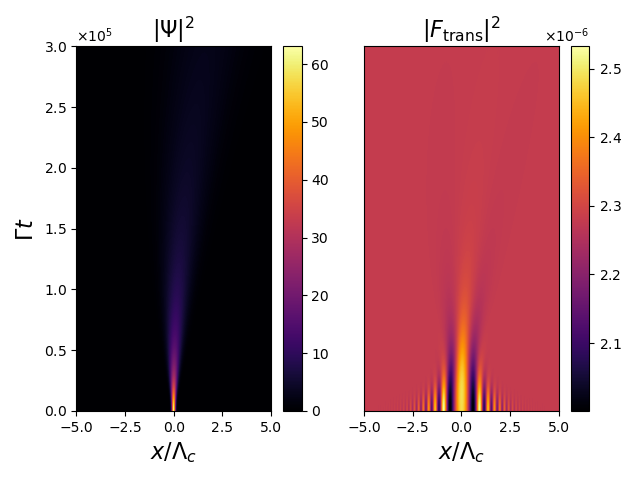}
		\caption{Evolution of  the BEC density, $|\Psi(\bar{x},\bar{t})|^2$,  and~optical intensity at the image plane, $|F_{\rm trans}(\bar{x},\bar{t})|^2$, calculated from the accelerating BEC--SMF model, Equation~(\ref{eqn:Ftr}), (\ref{eqn:B}) and (\ref{eqn:sch2}), with no mirror feedback ($R=0$). All other parameters are identical to those in Figure~\ref{fig:accelerating_red_droplet}.}
		\label{fig:nofeedback}
	\end{figure}

In order to demonstrate the significance of the mirror feedback for the droplet stability, Figure~\ref{fig:nofeedback} shows the evolution of the system when the mirror feedback is removed, i.e.,~mirror reflectivity $R=0$, with~all other parameters as  those in Figure~\ref{fig:accelerating_red_droplet}. It can be seen that in the absence of feedback, the droplet rapidly disperses and it is not possible to determine a well-defined trajectory from the similarly dispersing optical intensity distribution. This demonstrates that the mirror feedback is a critically important component of this system, which is required to maintain the soliton-like stability of the droplet and its corresponding optical~image.

\section{Conclusions}
We have demonstrated that the existence of optomechanical droplets in a BEC illuminated by a far off-resonant optical pump field and its retroreflection from a feedback mirror could form the basis of a method to measure the acceleration of a BEC via continuous monitoring of the position of the{droplet via} 
the optical intensity distribution. This differs from previous schemes involving acceleration sensing involving BECs in optical cavities, where the  acceleration was calculated via continuous measurements of the Bloch oscillation frequency~\cite{peden2009,venkatesh2009,goldwin2014,samoylova2015_1,samoylova2015_2,kessler2016} from the optical intensity~evolution.

It should be noted that although only the motion of single droplet structures has been presented here, stable multi-peak droplet structures can exist~\cite{walker2022} and display similar behaviour under acceleration, maintaining their structure as they propagate and providing a consistent optical intensity profile dependent on~detuning.

Possible extensions of the study presented here include increasing the interaction time by bouncing the BEC off a potential barrier and considering a spatially non-uniform acceleration of the~BEC.

\end{document}